\documentclass[twocolumn,showpacs,preprintnumbers,amsmath,amssymb]{revtex4}

\usepackage{graphicx}
\usepackage{dcolumn}
\usepackage{bm}

\begin{document}

\title{The rolling sphere and the quantum spin}

\author{Alberto G. Rojo}
\email{rojo@oakland.edu}
\affiliation{%
Department of Physics, Oakland University, Rochester, MI 48309.
}%
%
  \author{Anthony M. Bloch}
  \email{abloch@.umich.edu}
  \affiliation{%
Department of Mathematics, University of Michigan, Ann Arbor, MI
48109}

\date{\today}

  \begin{abstract}
We consider the problem of a sphere rolling of a curved surface and
solve it by mapping it to the precession of a spin $1/2$ in a
magnetic field of  variable magnitude and direction. The mapping can
be of pedagogical use in discussing both rolling and spin
precession, and in particular in understanding the emergence of
geometrical phases in classical problems.
  \end{abstract}
 \maketitle

\section{Introduction}

In this paper we consider a question similar to that posed in the
title of Ref. \cite{mont}: How much does a sphere rotate when
rolling on a curved surface? In Ref. \cite{mont},  the old problem
of the rotation of a torque free, non-spherical body is reanalyzed.
The angle of rotation  is identified to have two components, one
dynamical and one geometrical (the so called Berry phase),
independent of the time elapsed during the rotation. Here we
consider a related but different problem: a sphere is made to roll
without slipping on a given curve $\Gamma $ on a surface. The
question is, if the sphere completes a circuit, what is the rotation
matrix connecting the initial and final configuration of the sphere?
The problem we are considering is therefore a kinematic rather than
a dynamic one: the trajectory of the contact point of the sphere and
the surface is dictated externally and the rolling constraint is
imposed. We make contact with recent approaches that consider the
same problem \cite{levy1,johnson1} (but on a plane), in particular,
we address a nice question posed by Brockett and Dai
\cite{brockett}: a sphere lies on a table and is made to rotate by a
flat plane on top of it, parallel to the table. The question is: if
every point of the plane describes a circle, what is the trajectory
and motion of the sphere?

We treat the problem by exploiting  its isomorphism to the
precession of a spin 1/2 in a time-dependent magnetic field. In the
mapping, the arc length of the curve plays the role of time. For
rolling on a plane the magnitude of the magnetic field is $1/R$ with
$R$ the radius of the sphere, and the direction of the magnetic
field is that of the instantaneous angular velocity of the rolling
sphere. For a curved surface the normal curvature and the torsion of
the curve affect the value of the effective magnetic field. Closely
related to the present paper is the use of of the isomorphism
between classical dynamics and that of a spin $1/2$ by Berry and
Robbins in Ref. \cite{berryrobbins}, especially their classical view
of the Landau-Zener \cite{landauzener} problem.
 From a pedagogical perspective, the novel contribution of this paper is to use the isomorphism
 to discuss rolling spheres on an arbitrary surface.

The precession of a spin $1/2$ is widely treated in the literature
and one can borrow those results to acquire an intuition for the
rolling sphere. Conversely, since a rolling sphere is a tangible
physical problem, the present treatment can be useful pedagogically
in presenting spin precession, Berry's phases and it's classical
counterpart, Hannay's angle \cite{hannay}.


\section{Rolling on a plane and quantum precession}

Consider a sphere of radius $R$ rolling on a curve $\Gamma$ on a
plane. We define a local triad of unit vectors at the contact point
(the so called Darboux frame \cite{darboux}): the tangent
$\textbf{t}$ to $\Gamma$, the normal $\textbf{n}$ to the surface,
and $\textbf{u}=\textbf{n}\times \textbf{t}$,  the tangent normal.
For rolling on a plane $\textbf{n}$ is a constant vector, and the
velocity of the center of the sphere is along the tangent to the
curve. This situation will change for rolling on a curved surface,
but, as we will see, the general idea of the mapping to a precessing
spin is the same.

The translational velocity of the sphere is $\textbf{V}=\textbf{t}
V(t) $ and the rolling constraint  means that the instantaneous
velocity at the contact point is zero \cite{landau}:
\begin{eqnarray}
\overrightarrow{\omega}\times (\textbf{n}R)&=&\textbf{V} =\textbf{t}
V(t) \label{rollingc}
\end{eqnarray}
with $\overrightarrow{\omega}$ the angular velocity and $R$ the
radius of the sphere. This equation is nonintegrable and constitutes
a paradigmatic nonholonomic constraint \cite{bloch2003}.

Taking the cross product with $\textbf{n}$ on both sides of the
above equation we have

\begin{equation}
\overrightarrow{\omega} ={V(t)\over R} \textbf{n}\times
\textbf{t}\equiv  {V(t)\over R}  \textbf{u}.
\end{equation}

Notice that in the above equation we have used the ``no spin"
condition $\overrightarrow{\omega}\cdot \textbf{n}=0$, that is, we
are consider rolling without an instantaneous rotation along the
normal.

 The instantaneous velocity $\dot{\textbf{X}}$ of a point
of coordinate $\textbf{X}$ (with respect to the center of the
sphere) on the surface of the sphere is

\begin{equation}
\dot{\textbf{X}}=\overrightarrow{\omega} \times
\textbf{X}={V(t)\over R} \textbf{u}\times \textbf{X}. \label{xpp}
\end{equation}

Now we rewrite $V(t)=ds/dt$ where $s$ is the arc length of the curve
$\Gamma(t)$, and (\ref{xpp}) becomes
\begin{equation}
{d{\textbf{X}}\over d s}={\textbf{u}\over R}\times \textbf{X}.
\label{xpp1}
\end{equation}

If we regard $\textbf{X}=(x,y,z)$ as a magnetic moment, the  above
equation describes its precession in the presence of a magnetic
field $\textbf{B}=-{1\over R}(u_x,u_y,u_z)=-\overrightarrow{\omega}$
of constant magnitude $1/R$. The direction of $\textbf{B}$ is
$-\textbf{u}$, and varies varies with $s$, the arc length, which
plays the role of time. If the rolling is on a horizontal plane,
then $B_z$=0, but we keep this notation to make contact with the
rolling on an arbitrary surface.

There is an isomorphism between the rolling sphere written in this
way with a spin $1/2$ precessing in this magnetic field. This
isomorphism can be seen clearly if, (using
$\textbf{B}=-\overrightarrow{\omega}$)  we rewrite Equation
(\ref{xpp1}) in the form
\begin{equation}
{d\over d s}\left(\begin{array}{c}  x\\y\\z \end{array}\right) =
\left(\begin{array}{ccc}  0&B_z&-B_y\\
-B_z&0&B_x \\
B_y&-B_x&0
\end{array}\right)
\left(\begin{array}{c}  x\\y\\z\end{array}\right), \label{xmag1}
\end{equation}
which is the same as the following equation of motion for two
complex numbers $a$ and $b$ (we write $s$ instead of $t$ for time in
order to keep the analogy)
\begin{equation}
i{d\over d s}\left(\begin{array}{c}  a\\b \end{array}\right) =
-{1\over 2}\left(\begin{array}{cc}  B_z&B_x-iB_y\\
B_x+iB_y&-B_z \\
\end{array}\right)
\left(\begin{array}{c}  a\\b\end{array}\right), \label{eqspin}
\end{equation}
with the identification
\begin{eqnarray}
x&\equiv& ab^*+ba^* \nonumber
\\
y&\equiv& i\left(ab^*-ba^*\right) \nonumber \\
z&\equiv & aa^*-bb^*. \label{spin1}
\end{eqnarray}

The real numbers $(x,y,z)$ represent the coordinates of a point on
the surface of the sphere referred to a coordinate system fixed in
space (that is, not rotating), and whose origin is in the center of
the sphere. The above mapping is certainly possible because of the
$SU(2)-SO(3)$ isomorphism \cite{arfken}.

 Equation (\ref{eqspin}) is Schr\"odinger's equation for
the spinor $\chi=(a,b)$ in the presence of a magnetic field
$\textbf{B}$:
\begin{equation}
i{d\over ds}\chi=-\textbf{B}\cdot\textbf{S}\chi\equiv H \chi,
\end{equation}
where $\hbar=1$ and $H$ the Hamiltonian. Also, the vector
$\textbf{S}={1 \over 2}(\sigma_x,\sigma_y,\sigma_z)$ is the spin
operator, and $\sigma _i$ are Pauli's matrices. Notice that in this
mapping, the magnetic fields and the frequencies  have units of
inverse length,

Equation (\ref{spin1}) implies that we can extract the behavior of
the rolling sphere as a function of arc length by solving the motion
of a spin $1/2$ in a time-varying magnetic field. To our knowledge
the equivalence between the motion of rigid body and a two-level
system (a spin $1/2$), in the form of the mapping of Eq.
(\ref{spin1}) was first pointed out by Feynman, Vernon and Hellwarth
\cite{Feyn1} and later discussed several times \cite{ans}. Earlier,
Bloch \cite{bloch1} had derived the precession equation for the
density matrix of spin 1/2 and therefore the points $(x,y,z)$ that
result from the mapping from spinors are called the Bloch sphere.

 The pedagogical novelty of the present paper
 (an alternative title of which  could well have been ``The
rolling of the Bloch sphere") is to discuss the rolling using the
arc length as time and identifying the isomorphism between the
rolling sphere and the quantum spin in exactly solvable cases.


\section{Warmup: constant magnetic field}

Consider the simplest case of constant magnetic field. We choose
$\textbf{B}=B_0\hat{\textbf{k}}$, constant in the $+z$ direction.
This corresponds to the sphere rolling on a vertical plane. Eq
(\ref{eqspin}) becomes:
\begin{equation}
i{d\over d s}\left(\begin{array}{c}  a\\b \end{array}\right) =
-{1\over 2}\left(\begin{array}{cc}  B_0&0\\
0&-B_0 \\
\end{array}\right)
\left(\begin{array}{c}  a\\b\end{array}\right), \label{eqspin2}
\end{equation}

with solutions:
\begin{eqnarray}
\left(\begin{array}{c}  a(s)\\b(s) \end{array}\right) &=&
\left(\begin{array}{c}   e^{isB_0/2}a(0)\\
e^{-isB_0/2}b(0)\end{array}\right) \label{eqspin3}.
\end{eqnarray}

Replacing (\ref{eqspin3}) in (\ref{spin1}) we obtain:
\begin{eqnarray}
x(s) &=&x(0)\cos\left( {B_0s\over 2 }\right)+y(0)\sin\left(
{B_0s\over 2 }\right)\nonumber
\\
y(s) &=&y(0)\cos\left( {B_0s\over 2 }\right)-x(0)\sin\left(
{B_0s\over 2 }\right)\nonumber
\\
z(s)&=&z(0),
\end{eqnarray}
which means that the sphere is rotating clockwise around a constant
axis in the $z$ direction. This corresponds to
$\overrightarrow{\omega}$ in the $-z$ direction. In other words, a
constant magnetic field in the $z$ direction corresponds to the
sphere moving in a straight line in the $xy$ plane, rolling on a
vertical wall. The same situation applies if a constant field is
directed in any other orientation.

\section{The lollipop and the planar field}

\begin{figure}
\begin{center}
\vspace{-1.5cm}
\includegraphics[scale=.4]{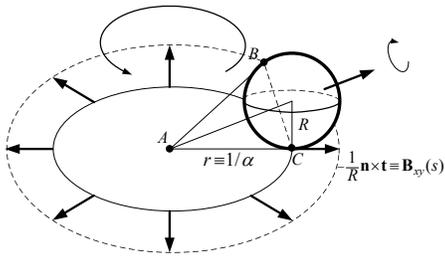}
\end{center}
 \vspace{-7.5cm}
\caption{The lollipop, or a sphere rolling counterclockwise on a
circle of radius $r$ corresponds to a spin $1/2$ precessing on a
magnetic field that rotates in the $xy$ plane.
 }
 \label{lolli}
\end{figure}

Consider a magnetic field varying on the $xy$ plane as
$\textbf{B}=B(\cos \alpha s, \sin \alpha s,0)$. This corresponds to
$\textbf{u}$ rotating with the same frequency in the same plane, and
the rolling problem becomes that of a sphere of radius $R=1/B$
rolling counterclockwise on a circle of radius $r=1/\alpha$ (see
Figure \ref{lolli}).

In turn, this corresponds to a time (or arc length) dependent
Hamiltonian $H=-\textbf{B}\cdot \textbf{S}$, which can be solved by
noting that

\begin{eqnarray}
\textbf{B}\cdot \textbf{S}&=&\left(\begin{array}{cc}  0&Be^{-i\alpha
s}
\\Be^{i\alpha s}&0\end{array}\right)
=U^*\left(\begin{array}{cc}  0&B\\
  B&0\end{array}\right)U,
  \label{tr1}
  \end{eqnarray}
  with
  \begin{eqnarray}
U&=&\left(\begin{array}{cc}  e^{i\alpha s/2}&0\\
0&e^{-i\alpha s/2}\end{array}\right). \label{tr2}
\end{eqnarray}

Substituting the above relations in (\ref{eqspin}) we obtain a time
independent equation for the coefficients $\tilde{\chi}(s)=(\tilde{
a},\tilde{b})= (e^{i\alpha s/2}a, e^{-i\alpha s/2}b)$

\begin{equation}
i{d\over d s}\left(\begin{array}{c} \tilde{ a}\\\tilde{b}
\end{array}\right) =
-{1\over 2}\left(\begin{array}{cc}  \alpha&B\\
B&-\alpha \\
\end{array}\right)\left(\begin{array}{c} \tilde{ a}\\
\tilde{b}
\end{array}\right)\equiv \tilde{ H}\left(\begin{array}{c} \tilde{ a}\\
\tilde{b}
\end{array}\right) .
\label{eqspintr}
\end{equation}

Transformations (\ref{tr1}) and (\ref{tr2}) correspond to
transforming to a frame that rotates with angular velocity $\alpha $
\cite{slichter}. When transforming to the rotating frame, the
angular velocity acquires a component $\alpha=1/r$ in the  $z$
direction and the frequency of rotation in the rotating frame is

\begin{equation}\Omega=\sqrt {B^2+\alpha ^2}= {1\over rR}\sqrt{r^2+R^2}
\label{omegalol}
\end{equation}

This can be seen in the spinor language by noting that, since
$\tilde{H}$ in Eq. (\ref{eqspintr}) is time-independent , the
solutions are
\begin{eqnarray}
\tilde{\chi}(s)
&=&e^{{i\over 2}s\left(\begin{array}{cc}  \alpha&B\\
B&-\alpha \\
\end{array}\right)}
\tilde{\chi}(0) \nonumber \\&=& \left[\cos\left({\Omega s /
2}\right)+ i\vec{\sigma}\cdot \textbf{m}\sin \left({\Omega s /
2}\right) \right] \tilde{\chi}(0), \label{spinrot}
\end{eqnarray}
with $\pm\Omega=\sqrt{B^2 +\alpha^2}$ the eigenvalues of $\tilde{H}$
and $\textbf{m}$ a unit vector in the direction $\alpha/B=R/r$.
Equation (\ref{spinrot}) describes a rotation at a rate $\Omega$
with respect to an axis in the direction of the ``stick" of the
lollipop (the direction joining $A$ to the center of the sphere (see
Fig. (\ref{lolli})). Notice that solving for the evolution by
exponentiating $\tilde{H}$ is possible because $\tilde{H}$ does not
depend on $s$. If there is an $s$-dependence and the matrices
$\tilde{H}$ at different $s$ do not commute the solution is a ``time
ordered" exponential that in general is not exactly solvable.

After the lollipop completes a circle, the angle $\delta$ of
rotation is

\begin{equation}
\delta= {2\pi \over \alpha}\Omega=2\pi\sqrt{1+\left({r\over
R}\right)^2}. \label{delta1}
\end{equation}

Notice that, when $R\ll r$ the angle of rotation is $\delta= 2\pi
r/R$, corresponding to rolling in a line of length equal to the
perimeter of the circle.

In anticipation of the next section we mention than in this case,
since the rolling is on the plane, there is no geometric phase. When
the rolling is on a curved surface the situation changes. Notice
that we are using the term ``geometric phase" in its relation with
the spin problem in the adiabatic approximation. This phase is
different from the nonholonomy when the sphere of arbitrary radius
describes a loop.

We see that, after traveling on a circle the sphere is rotated by
$2\pi \Omega/\alpha$ with respect to an axis tilted with respect to
the plane; this is the nonholonomy treated in \cite{johnson1} and
\cite{iwai}.

When the sphere rolls on a plane, and on a circle of radius $r$ much
larger than its radius $R$, it comes back rotated around an axis
that lies on the plane, by an angle given only by the dynamical
phase. The extra term that originates in the curvature of the
surface is what we call the geometric phase.

The angle of rotation $\delta$ (of both the spin and the lollipop)
has a simple geometric interpretation: when the lollipop rolls, the
point of contact $C$ moves on the circular rim of the cone $ABC$
(see Figure \ref{lolli}). At the same time, the point $C$ ``paints"
on the sphere a circle of diameter $BC=2rR/\sqrt{r^2+R^2}$. (This is
easily calculated with simple geometrical considerations from Figure
\ref{lolli}.) This means that after a revolution of length $2\pi r$
the angle rotated is $2\pi r /(BC/2)$ from which Eq. (\ref{delta1})
follows immediately.

At this point we consider Brockett's question mentioned in the
Introduction. Notice first that, as the sphere rolls on a circle,
the velocity at the top of the sphere is twice the velocity
$\textbf{V}$ at the center of the sphere. Since each point of the
plane on top of the sphere describes a circle of radius $R_1$, the
velocity $\textbf{V}_P$ of the plane also describes a circle.
Therefore, since the sphere has a rolling condition with the upper
plane, then $\textbf{V}_P=2\textbf{V}$, meaning that, as the plane
describes a circle of radius $R_1$ the sphere describes a circle of
radius $R_1/2$.

We showed this with a nice classroom demo: on a piece of paper draw
a circle of radius 5 inches (twice that of a tennis ball). Orient
the label of the tennis ball at 45 degrees with the vertical (the
sphere is going to roll on a circle of radius $r=R$, and therefore
the axis of rotation is going to be at 45 degrees and the precession
frequency will be, from (\ref{omegalol}), $\sqrt{2}$). Paint a mark
on a transparent glass, which in turn will serve as the upper plane.
Also mark three points on the circle separated by $\beta=127$
degrees ($\pi/\sqrt{2}$). Looking through the glass, guide the mark
on the glass over the circle on the paper, and notice that, each
time the glass rotates by $\beta$, the tennis ball rotates by $\pi$
with respect to a moving axis at 45 degrees.

Notice also that for $s=2\pi/\alpha$ the spinor $\chi $ changes sign
due to the $1/2$ factor in the transformation. Nevertheless,  since
the mapping of (\ref{spin1}) is quadratic in $a$ and $b$, changing
their signs corresponds to the same values $(x,y,z)$ for the
orientations. More specifically, the quantities $a$ and $b$
determine univocally $x, y$ and $z$, but the reverse is not valid:
the quantum evolution determines univocally the classical evolution
but there is some ambiguity in going from the classical to the
quantum case. For example if we perform the ``gauge transformation"
$(a,b) \rightarrow e^{i\phi(s)}(a,b)$ the mapping to the
$\textbf{X}$ coordinate remains unchanged.

We will come back to this point in the next sections when we discuss
the geometric phase for rolling.

\section{rolling on a curved surface}

\begin{figure}
\vspace{-1.5cm}
\includegraphics[scale=.4]{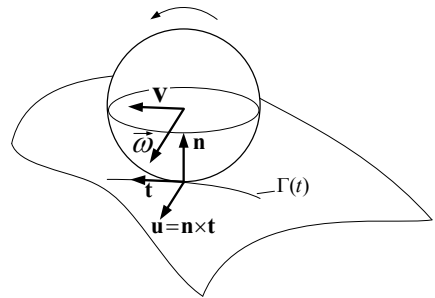}
 \hspace{-1.cm}
 \vspace{-6.cm}
\caption{{ Sphere rolling along a curve $\Gamma$ of zero torsion
(meaning that the velocity of the center of the sphere is parallel
to the tangent of the curve at the contact point). }
 }
 \label{harmfig1}
\end{figure}

In this section we extend the treatment of rolling on a plane to
rolling on a curved surface (See Figure \ref{harmfig1}). If we call
$\textbf{X}_P$ the coordinate of the contact point, the coordinate
$\textbf{X}_c$ of the center of the sphere is:
\begin{equation}
\textbf{X}_c=\textbf{X}_P + R\textbf{n},
\end{equation}
and its velocity is given by
\begin{eqnarray}
\dot{\textbf{X}}_c&=&\dot{\textbf{X}}_P + R\dot{\textbf{n}},
\nonumber
\\
&=& \left( \textbf{t} + R {d\textbf{n}\over ds}\right) {ds\over dt}.
\label{dotxc}
\end{eqnarray}

The rolling condition is that the velocity of a point of the sphere
in contact with the surface is zero (See Eq.(\ref{rollingc})):

\begin{equation}
\overrightarrow{\omega}\times (\textbf{n}R)=\dot{\textbf{X}}_c.
\end{equation}

Again, taking the cross product with $\textbf{n}$ on both sides of
the equation above we obtain

\begin{equation}
\overrightarrow{\omega}={1\over R} \textbf{n}\times
\dot{\textbf{X}}_c. \label{omcurve}
\end{equation}

We now replace (\ref{dotxc}) in (\ref{omcurve}), and use the fact
that, for a curved surface, the variation of the normal is given by
\begin{equation}
{d\textbf{n}\over ds}=-\kappa_n \textbf{t}- \tau_r \textbf{u},
\end{equation}
with $\kappa_n$ the normal curvature and $\tau_r$ the torsion of the
curve, both evaluated at the contact point. We obtain

\begin{equation}
\overrightarrow{\omega}=\left[{1\over R} \left(
1-\kappa_nR\right)\textbf{u}+ {\tau _r }\textbf{t}\right]{ds\over
dt}. \label{omcurve2}
\end{equation}

The discussion for the planar case extends to the curved surface,
and the rolling of the sphere is equivalent to a spin $1/2$
precessing on a magnetic field $\textbf{B}(s)$ given by
\begin{equation}
\textbf{B}(s)=-\left[{1\over R} \left( 1-\kappa_nR\right)\textbf{u}+
{\tau _r }\textbf{t}\right], \label{omcurve3}
\end{equation}
with the arc length $s$ playing the role of time. In the following
section, as an example of this formulation we consider rolling on a
spherical surface.

\section{sphere rolling on a spherical surface}

\begin{figure}
\begin{center}
\vspace{.cm}
\includegraphics[scale=.3]{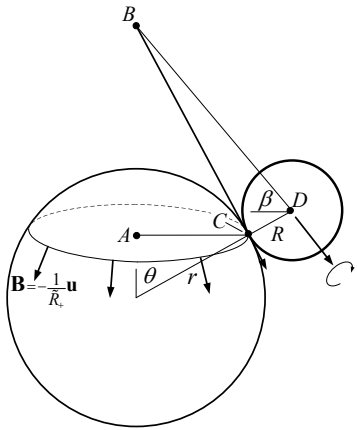}
\end{center}
 \vspace{-1.5cm}
\caption{Sphere rolling on a sphere.
 }
 \label{Fig3}
\end{figure}

In this section we consider a  sphere of radius $R$ rolling on a
second sphere of radius $r$. The rolling line will be a parallel of
latitude $\pi/2 -\theta$  (see Figure \ref{Fig3}). This means that
the normal curvature is constant $1/r$, and also that the torsion is
zero. The magnetic field for the corresponding spin problem is
therefore:

\begin{equation}
\textbf{B}(s)=-\left[{1\over R} \left( 1\pm {R\over
r}\right)\textbf{u}\right]=-{1\over \widetilde{R}_{\pm}}\textbf{u},
\label{bsphere}
\end{equation}
with $\widetilde{R}_{\pm}=rR/(r\pm R)$  a reduced radius  and the
plus and minus signs refer to the rolling outside and inside of the
sphere of radius $r$ respectively.

For a sphere rolling on a parallel,  the instantaneous angular
velocity (and the magnetic field) describes a cone forming an angle
$\theta$ with the vertical. The total arc length of the parallel is
$r \sin \theta $ meaning that the vector $\textbf{u}$ rotates with
angular frequency $\alpha$ given by  $\alpha = 1/(r\sin\theta)$. The
corresponding magnetic field is therefore
\begin{equation}
\textbf{B}(s)=(B_x,B_y,B_z)={1\over \widetilde{R}_{\pm}}(\cos\theta
\cos \alpha s,\cos \theta \sin \alpha s, -\sin \theta)
\end{equation}
with the term $\textbf{B}\cdot \textbf{S}$ in the corresponding
Hamiltonian given in this case by
\begin{eqnarray}
\textbf{B}\cdot \textbf{S}&=&{1\over 2}{1\over
\widetilde{R}_{\pm}}\left(\begin{array}{cc} -\sin \theta &\cos
\theta e^{-i\alpha s}
\\\cos \theta e^{i\alpha s}&\sin \theta\end{array}\right).
 \label{bssphere}
  \end{eqnarray}

This again is an exactly solvable Hamiltonian that was first studied
by Rabi

  Using the same transformation matrix of Eq. (\ref{tr2}) the above
  Hamiltonian can be rendered time independent. We write it in the
  following form


\begin{eqnarray}
\tilde {H}&=&-{1\over 2}\left(\begin{array}{cc}- B_0 \sin \theta +
\alpha  & B_0 {\cos \theta }
\\B_0{\cos \theta} &B_0{\sin \theta }- \alpha
\end{array}\right),
 \label{trh}
  \end{eqnarray}
with $B_0=1/\widetilde{R}_{\pm}$.


 The eigenvalues of $\tilde{H}$
are

 \begin{eqnarray}
  \Omega _{\pm}
&=&{1\over \widetilde{R}_{\pm}}\sqrt{1-{2\widetilde{R}_{\pm}\over
r}+ \left({\widetilde{R}_{\pm}\over r\sin\theta}\right)^2 },
\label{omega2s}
  \end{eqnarray}
with the spinor
  precessing,
   in
  the rotating frame, around an axis that forms an angle $\beta $
  (see Figure \ref{Fig3})
  with the $xy$ plane, with
  \begin{equation}
  \tan \beta = \tan \theta - {R\over r+R }{1\over \sin \theta \cos
  \theta}
  \label{newrot}
  \end{equation}

The second term in (\ref{newrot}) reflects the fact that the small
sphere rotates instantaneously on the tangent plane that contains
$BC$ (see Figure \ref{Fig3}).
 Equation (\ref{newrot}) can be easily
derived by simple geometric considerations from Figure (\ref{Fig3}).

After a complete revolution the angle or rotation $\delta $ is
\begin{equation}
\delta_{\pm} = 2\pi r \sin \theta \Omega_{\pm}.
\end{equation}

After a little algebra we obtain
 \begin{eqnarray}
  \delta _{+} &=& 2\pi \cos \theta \sqrt{1+\left({r\tan  \theta\over R} \right)^2},\nonumber
\\
 \delta_{-} &=&2\pi \sqrt{1+\sin^2\theta {(r-3R)(r-R)\over R^2}}.
\end{eqnarray}

Notice that, if we compare with the rotation in a plane from Eq.
(\ref{omegalol}), the ``outer" rotation corresponds to rolling on a
circle of radius equal to that of the unfolded cone tangent to the
parallel (See Fig. \ref{Fig3}). The angle of rotation along that
circle is not $2\pi$ but $2\pi \cos \theta$.
This geometric factor is the same
that appears in Foucault's pendulum and in Berry's phase for a spin
precessing on a cone (we will come back to this point below). Also,
notice that when $r=R$ the angle of rotation is always $2\pi$
independent of latitude.

The ``inner" roll has the interesting feature that, when $r=3R$ the
angle of rotation is $2\pi$ regardless of latitude. This amusing
feature can be verified easily at the equator: roll a penny inside
of a circle of radius three times the radius of the penny and verify
that the penny completes a full rotation in the rotating frame (and
of course two full rotations in the lab frame).

We finish this section with a discussion of the differences and
similarities between the Berry phase for a precessing spin $1/2$ in
the adiabatic approximation and the rolling of two spheres.

The Hamiltonian for a spin in a magnetic field that precesses along
the $z$ axis at frequency $\alpha $ is given by (\ref{trh}), where
in principle $\alpha$ and $B_0$ are independent parameters. If
$\alpha \ll B_0$ (the adiabatic approximation) the eigenvalues
(eigenfrequencies) of $\widetilde{H}$ are

\begin{equation}
\Omega\simeq \sqrt{B_0^2-2\alpha B_0 \sin \theta} \simeq B_0 -\alpha
\sin \theta.
\end{equation}

After a period of time $2\pi /\alpha$ the change $\Delta \phi $ in
the phase of the spin is
\begin{equation}
\Delta \phi = 2\pi {B_0\over \alpha} - 2\pi \sin \theta.
\label{berry1}
\end{equation}

The first therm is the dynamical phase and the second is a purely
geometrical one, independent of the parameters $B_0$ and $\alpha$,
and give by (half) the solid angle described by the field.

For the rolling sphere we can also study an ``adiabatic
approximation" since $\alpha \ll B_0$  corresponds to $r\gg R$. In
other words, in general the adiabatic approximation will correspond
to the radius of the rolling sphere much smaller than the radius of
curvature of the surface. On the other hand, in contrast with the
spin case, the frequency of rotation $\alpha =1/r\sin \theta $
``knows" about the latitude and the curvature. So we expect some
differences and some similarities. Replacing the values of
$B_0=1/\widetilde{R}_{\pm}\equiv 1/R \pm 1/r$ in (\ref{berry1}) we
obtain the angle of rotation of the sphere in each case (in the
adiabatic approximation)
\begin{eqnarray}
\Delta \phi_{\pm} &=& 2\pi r \sin \theta \left( {1\over R} \pm
{1\over r} \right) - 2\pi \sin \theta. \nonumber \\
&=& \left\{
\begin{array} {cc}
2\pi {r \sin \theta \over R} & {\rm{ (Outer \;\; rolling)}},
\\
2\pi {r \sin \theta \over R} - 4\pi \sin \theta & {\rm{ (Inner \;\;
rolling)}}
\end{array}
\right.
 \label{berry2}
\end{eqnarray}

The above interplay of curvatures for inner and outer rolling is a
special case of more general treatments of kinematics of rolling and
is discussed in Ref. \cite{montana}.

From Eq. (\ref{berry2}), we see that in the outer rolling case there
is no Berry phase, something we could have expected because of the
analogy with the rolling on a flat plane. The angle of rotation is
in this case given simply by the rotation on a straight line of
length equal to the perimeter of the parallel. However, for the
inner rolling we indeed have a geometric phase twice as big as that
of the spin $1/2$. Our treatment is a nice example of the appearance
of a geometric phase in a classical system, originally discussed by
Hannay \cite{hannay}.



In the next section we discuss the general  connection between
rolling and
 the
Berry phase for spins in the adiabatic approximation.

\section{The adiabatic approximation and rolling on a curved surface}

In this section we compare the equivalence between the adiabatic
approximation for a spin precessing in a magnetic field that changes
direction at a slow rate and rolling on a surface. In the spin case,
the dimensionless parameter controlling the approximation is the
ratio of the instantaneous frequency (proportional to the
instantaneous magnitude of the field) with the rate at which it's
direction is changing.

In the rolling case the instantaneous frequency corresponds to the
magnitude of  $\textbf{B}(s)$ and the rate of change in its
direction is related to the normal curvature and to the curve's
torsion.

In the adiabatic approximation for spins \cite{sakurai}, one works
in an ``instantaneous" basis, treating first $s$ (time) as a
parameter and solving the eigenvalue equation as though the problem
were static:

\begin{equation}
H(s)\chi(s)=\Omega(s)\chi(s).
\end{equation}

Then the general solution is written as linear combinations of the
instantaneous eigenstates. As a result, in the adiabatic
approximation, the spinor at time $s$ is given by

\begin{equation}
\chi(s)=e^{i\gamma(s)}e^{i\int_0^s ds' \Omega(s')}\chi(0).
\end{equation}

The argument of the second exponential above represents the dynamic
phase, which involves the integral of the following angular
frequency:
\begin{eqnarray} \Omega(s)&=& |\textbf{B}(s)|={1\over R}
\sqrt{\left[1-\kappa_n(s)R\right]^2+[\tau(s)R]^2} \nonumber \\
&\simeq& {1\over R} - \kappa_n(s)
\end{eqnarray}

This can be seen, for example from Equation (\ref{bssphere}): the
eigenvalues of $\textbf{B}\cdot \textbf{S}$ with $s$ treated as a
parameter are $\pm |\textbf{B}(s)|$.

The (instantaneous) direction of the field is in the direction
$\textbf{u}_B$ given by

\begin{equation}
\textbf{u}_B={\textbf{B}(s)\over |\textbf{B}(s)|}= -{ \left(
1-\kappa_nR\right)\textbf{u}+ {\tau _r }\textbf{t} \over
\sqrt{\left( 1-\kappa_nR\right)^2+ \tau _r^2}}
\end{equation}

In general, the eigenvalues of a Pauli matrix in an arbitrary
direction $\textbf{u}_B\cdot \vec{\sigma}$ given by the unit vector
$\textbf{u}_B=(u_x, u_y, u_z)$ are $\pm 1$. This is verified by
noting that (defining $u_x+iu_y=\rho e^{i\phi}$)
\begin{equation}
\left(\textbf{u}_B\cdot
\vec{\sigma}\right)\chi_{\pm}(\textbf{u}_B)=\left(\begin{array}{cc}
u_z& \rho e^{-i\phi}\\ \rho e^{i\phi}&- u_z
\end{array}\right)\chi_{\pm}(\textbf{u}_B)=\pm\chi_{\pm}(\textbf{u}_B),
\end{equation}
with $\chi_{\pm}(\textbf{u}_B) =( 1, \pm (1-u_z)e^{\pm
i\phi}/\rho)$. Notice that the dependence of $\chi$ on $s$ is
through the orientation of $\textbf{u}$.

The first term, the geometric phase $\gamma$, is the Berry phase,
and is given by

\begin{equation}
\gamma(s)= \int_0^s ds' \chi(\textbf{u}_B(s')) ^{\dagger}{d \over
ds'}\chi (\textbf{u}_B(s')).
\end{equation}

If the rolling describes a complete circuit, $\gamma$ measures the
solid angle described by $ \textbf{u}_B$. This can be seen
explicitly as follows. If we express $\textbf{u}_B$ in polar
coordinates $\textbf{u}_B=(\sin \theta \cos \phi, \sin \theta \cos
\phi, \cos \theta)$ then the spinor in that direction is:

\begin{equation}\chi(\textbf{u}_B(s))=\left( \begin{array}{c}
\cos {\theta (s)\over 2}e^{-i\phi(s)/2}\\\sin {\theta(s)\over
2}e^{i\phi(s)/2}
\end{array}\right).
\end{equation}

This means that $ \chi ^{\dagger}{d \over ds'}\chi =-{1\over 2} \cos
{\theta } {d\phi \over ds'} $, and the integral over a closed
circuit $\Gamma $ can be written as
\begin{equation}
\gamma_{\Gamma}= -{1\over 2} \oint _{\Gamma} \textbf{A}\cdot
d\textbf{l},
\end{equation}
with $\textbf{A}={\cos{\theta}\over \sin{\theta}}
\textbf{u}_{\phi}$. Since $\nabla \times \textbf{A}=-\textbf{u}_r$,
using Stokes theorem, the line integral of $\textbf{A}$ is the flux
of a monopole in the origin, giving the solid angle \cite{holstein}.

Notice that
 this solid is traced not by the normal to the surface but by
 $ \textbf{u}_B$. This means that the solid angle measures
 a combination of the normal curvature and the torsion of the curve.
  In contrast, the solid angle traced by the normal measures the
geodesic curvature \cite{levy2}.  In summary, we have shown that,
when a sphere of radius $R$ rolls on a surface of local radius of
curvature and inverse torsion much larger than $R$, the angle of
rotation $\delta$ in a closed curve of length $L$ is given by

\begin{equation}
\delta ={L\over R} -\int _0^L ds \kappa_n(s)- S, \label{original}
\end{equation}
with $S$ the solid angle traced by $\textbf{u}_B$. Note that, if we
specify this result to the sphere  rolling on the parallel of a
sphere of radius $r$, we have $L=2\pi r \sin \theta$, $\kappa_n=\pm
1/r$ (for inner and outer rolling respectively) and $S=2\pi \sin
\theta$. Replacing these in Eq. (\ref{original}) we obtain the
result of Eq. (\ref{berry2}) as expected.

\section{acknowledgments}
We thank Michael V. Berry for useful comments on the manuscript and
for pointing us to Ref \cite{berryrobbins}. We thank Roger Brockett
and Paul R. Berman for interesting remarks. A.G.R thanks the
Research Corporation, and A.M.B. thanks the National Science
Foundation for support.

\end{document}